\documentclass[namedreferences]{solarphysics}
\usepackage[optionalrh]{spr-sola-addons} 
\usepackage{graphicx}        
\usepackage{color}           
\usepackage{url}             
\usepackage[cp1250]{inputenc}

\newcommand{\etal}{{\it et al. }}

\begin{document}
\begin{article}
\begin{opening}

\title{Fast magnetoacoustic waves in a~fan structure above the coronal magnetic null point}

\author{H. \surname{M\'esz\'arosov\'a}$^{1,2}$\sep
        J. \surname{Dud\'ik}$^{2,3}$\sep\\
        M. \surname{Karlick\'y}$^{2}$\sep
        F. R. H. \surname{Madsen}$^{1}$\sep\\
        H. S. \surname{Sawant}$^{1}$\sep }

\institute{$^{1}$National Space Research Institute (INPE),
                 Ave. dos Astronautas 1758, 1221-0000 S$\tilde{a}$o Jos\'e dos Campos, SP,
                 Brazil, \email{frhmadsen@gmail.com, sawant@das.inpe.br}\\
           $^{2}$Astronomical Institute of the Academy of Sciences,
                 CZ-25165 Ond\v{r}ejov, Czech Republic,
                 \email{hana@asu.cas.cz, karlicky@asu.cas.cz} \\
           $^{3}$Dept. of Astronomy, Physics of the Earth and Meteorology,
         Faculty of Mathematics, Physics and Informatics, Comenius University,
                 SK-842 48 Bratislava, Slovak Republic,
                 \email{dudik@fmph.uniba.sk} \\}

\runningauthor{M\'esz\'arosov\'a et al.}

\runningtitle{Fast magnetoacoustic waves in a~fan structure}

\date{Received ; accepted }

\begin{abstract}
We analyze the 26~November~2005 solar radio event observed interferometrically at
frequencies of 244 and 611\,MHz by the \textit{Giant Metrewave Radio Telescope} (GMRT) in
Pune, India. These observations are used to make interferometric maps of the event at
both frequencies with the time cadence of 1\,s from 06:50 to 07:12\,UT. These maps reveal
several radio sources. The light curves of these sources show that only two sources at
244\,MHz and 611\,MHz are well correlated in time. The EUV flare is more localized with
flare loops located rather away from the radio sources. Using the SoHO/MDI observations
and potential magnetic field extrapolation we demonstrate that both the correlated
sources are located in the fan structure of magnetic field lines starting from a~coronal
magnetic null point. Wavelet analysis of the light curves of the radio sources detects
tadpoles with periods in the range $P$~=~10--83\,s. These wavelet tadpoles indicate the
presence of fast magnetoacoustic waves that propagate in the fan structure of the coronal
magnetic null point. We estimate the plasma parameters in the studied radio sources and
find them consistent with the presented scenario involving the coronal magnetic null
point.
\end{abstract}

\end{opening}

\section{Introduction}
\label{Sect:1}
It is commonly believed that the primary energy release regions in solar flares are
located in the low corona \cite{Aschwanden05}. Radio spectral and imaging observations in
the decimetric and metric wavelength ranges in combination with magnetic field
extrapolations are considered to be very promising tools for a~study of these processes.
Unfortunately, imaging observations of solar flares in the decimetric range are very
rare. In the metric range, such observations are commonly done by the Nan\c{c}ay
radioheliograph \cite{Kerdraon97}.

To successfully combine radio observations and magnetic extrapolations, the radio
observations have to include sufficiently precise positional information. Such
a~combination was done \textit{e.g.} by \inlinecite{Trottet06}, where a~detailed analysis
of radio spectral and imaging observations in the 10--4500\,MHz range was presented for
the 5~November~1998 flare.  \inlinecite{Subramanian07} have studied a~post-flare source
imaged at 1060\,MHz to calculate the power budget for the efficiency of the plasma
emission mechanism in a~post-flare decimetric continuum source. \inlinecite{Aurass07}
have analyzed the topology of the potential coronal magnetic field near the source site
of the meter-decimeter radio continuum to find that this radio source occurs near the
contact of three separatrixes between magnetic flux cells. \inlinecite{Aurass11} have
examined meter-decimeter dynamic radio spectra and imaging with longitudinal magnetic
field magnetograms to describe meter-wave sources. \inlinecite{Chen11} have used an
interferometric dm-radio observation and nonlinear force-free field extrapolation to
explore the zebra pattern source in relation to the magnetic field configuration.
\inlinecite{Zuccarello11} investigated the morphological and magnetic evolution of an
active region before and during an X3.8 long duration event. They found that coronal
magnetic null points played an important role in this flare.

A~comprehensive review of magnetic fields and magnetic reconnection theory as well as
observational findings was provided by \inlinecite{Aschwanden05}. The topological methods
for the analysis of magnetic fields were reviewed in \inlinecite{Longcope05} and the
3D~null point reconnection regimes in \inlinecite{Priest09}.  \inlinecite{McLaughlin11}
have presented a~review of the theoretical studies of the MHD waves in the vicinity of
the magnetic null points. Furthermore, \inlinecite{Afana12} have studied analytically the
propagation of a~fast-mode magnetohydrodynamic wave near a~2D magnetic null point. Using
the nonlinear geometrical acoustic method they have found complex behavior of these waves
in the vicinity of this point. In spite of the wealth of theoretical work presented in
these papers, the authors concluded that there is still no clear observational evidence
for the presence of MHD waves near null points of the magnetic field. We note that this
is also in spite the fact that MHD waves are commonly observed in the solar corona
\cite{DeMoortel00,Kliem02,Harrison02,Tomczyk07,Ofman08,DeMoortel09,Marsh09,Marsh11}.

%
%
\begin{figure}[]
\begin{center}
\includegraphics[scale=0.60]{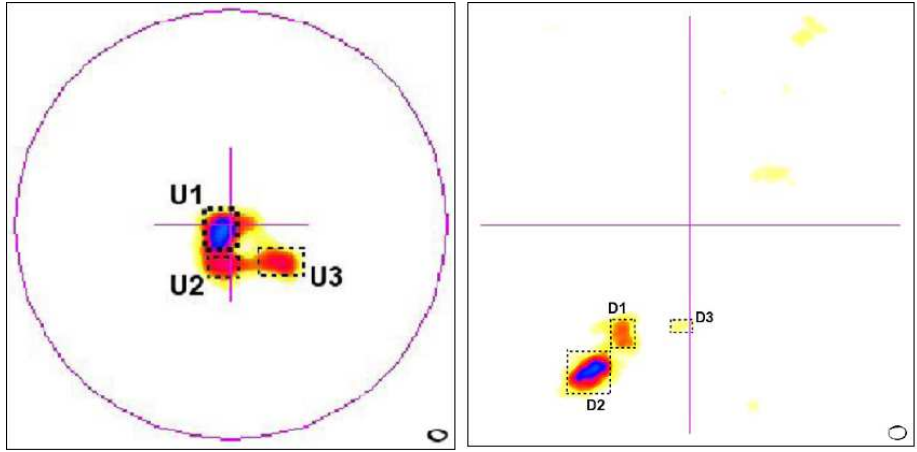}
\caption{Solar maps with main sources associated with the 26~November~2005 radio event
         observed at 7:01:30\,UT by the GMRT instrument. Regions U1--U3 and D1--D3 were
         identified as the main radio sources at 244 and 611\,MHz, respectively. The magenta
         circle in the left panel has the diameter of 32\,arc\,min and indicates an
         approximate position and size of the visible solar disk. Disk centers are
         shown by a~cross. The cross has dimensions of 400'' in both directions to indicate
         the scale in the maps. Synthesized beam dimensions giving the error in GMRT positions
         are represented by the small ovals shown on the bottom right corners.}
\label{figure1}
\end{center}
\end{figure}

Since both the standing and propagating magnetoacoustic waves modulate the
plasma density and the magnetic field in the radio source \cite{Aschwanden05},
some modulation of the radio emission by both these waves can be expected.

\citeauthor{Roberts83} (\citeyear{Roberts83,Roberts84}) studied impulsively generated
fast magnetoacoustic waves trapped in a~structure with enhanced density (\textit{e.g.}
loop). They showed that these propagating waves exhibit both periodic and quasi-periodic
phases. \inlinecite{Nakariakov04} numerically modelled impulsively generated fast
magnetoacoustic wave trains and showed that the quasi-periodicity is a~consequence of the
dispersion of the guided modes. Using wavelet analysis, these authors found that typical
wavelet spectrum of such fast magnetoacoustic wave trains is a~tadpole consisting of
a~broadband head preceded by a~narrowband tail.

The tadpoles as characteristic wavelet signatures of fast magnetoacoustic wave trains
were observed in solar eclipse data \cite{Katsiyannis03} as well as in radio spectra of
decimetric gyrosynchrotron emission \cite{Meszarosova09a}, and also in decimetric plasma
emission \cite{Meszarosova09b}. While the tadpoles in the gyrosynchrotron emission was
detected simultaneously at all radio frequencies, the tadpoles in the plasma emission
drifted towards low frequencies. This type of ``drifting tadpoles'' was studied in
details in \inlinecite{Meszarosova11} in the radio dynamical spectrum with fibers.

The observed parameters of fast magnetoacoustic waves reflect properties of the plasma in
the waveguides  where these waves are propagating. Therefore, one could use observed
waves and their wavelet tadpoles as a~potentially useful diagnostic tool
\cite{Jelinek09,Jelinek10} for determining physical conditions in these waveguides,
(\textit{e.g.} loops or current sheets). \inlinecite{Karlicky11} compared parameters of
wavelet tadpoles detected in the radio dynamical spectra with narrowband spikes to those
computed in the model with the Harris current sheet. Based on this comparison the authors
proposed that the spikes are generated by driven coalescence and fragmentation processes
in turbulent reconnection outflows. We note here that, in general, flare current sheets
can be formed not only near magnetic null points, but also \textit{e.g.} between
interacting magnetic loops. \inlinecite{Jelinek12} numerically studied impulsively
generated magnetoacoustic waves for the Harris current sheet and a density slab. In both
cases they find that wave trains were generated and propagated in a~similar way for
similar geometrical and plasma parameters.

In this paper, we analyze a~rare decimetric imaging observation of the 26~November~2005
solar flare made by the \textit{Giant Metrewave Radio Telescope} (GMRT). Combing the results of
this analysis with the magnetic field extrapolation (Section~\ref{Sect:3}), we present
a~scenario of this radio event. For the first time, we detected the magnetoacoustic waves
in the radio sources (Section~\ref{Sect:4}) located in the fan of magnetic field lines
connected with a~coronal null point. The basic plasma parameters in the radio sources are
estimated and the results are discussed (Section~\ref{Sect:5}).

%
%
\begin{figure}[]
\begin{center}
\includegraphics[scale=0.65]{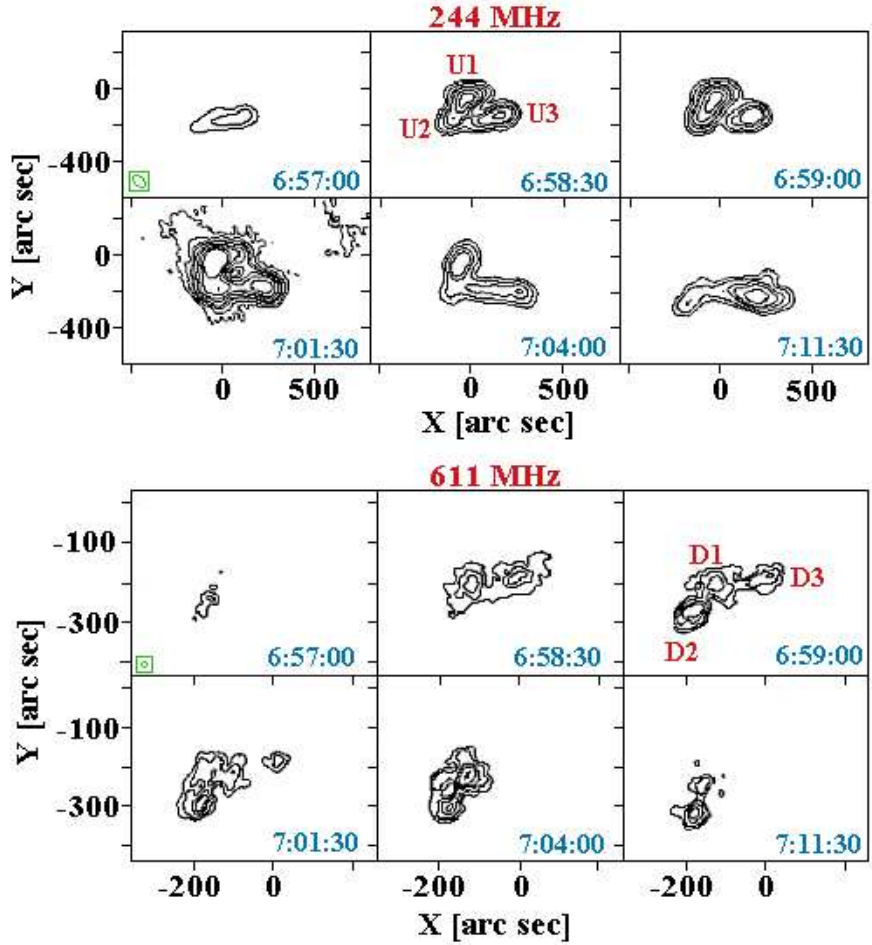}
\caption{Selected contour maps showing the time evolution of the emission sources at 244\,MHz
         (upper panel) and 611\,MHz (bottom panel). The times (UT) are in blue.
         Synthesized beam dimensions representing the error in GMRT positions are shown as
         small green circles on the bottom left corners of maps at 6:57\,UT.
         Upper panel: Positions of the sources U1--U3 (in red) are indicated on the map at 6:58:30\,UT.
         Bottom panel: Positions of the sources D1--D3 (in red) are indicated on the map at 6:59\,UT.}
\label{figure2}
\end{center}
\end{figure}

%
%
\begin{table}[]
\caption[]{Time evolution of the individual GMRT sources.}
\label{table1}
\begin{tabular}{cccc}
\hline
\hline
Source & Start time & Time of max & End time \\
       &  $[$UT]      &   [UT]     &  [UT]    \\
\hline
U1     & 6:58:04    & 7:01:30      & 7:04:03  \\
U2     & 6:57:59    & 7:01:30      &          \\
U3     & 6:58:20    & 7:06:45      &          \\
D1     & 6:58:05    & 7:03:52      & 7:05:11  \\
D2     & 6:58:54    & 6:59:33      & 7:04:00  \\
D3     & 6:58:13    & 6:58:43      & 7:01:59  \\
\hline
\end{tabular}
\end{table}

%
%
\section{Observations and data analysis}
\label{Sect:2}
The B8.9~solar flare occurred on 26~November~2005 in the active regions NOAA AR~10824 and
10825 located near the disk center. The flare lasted from approximately 06:31 to
07:49\,UT with the GOES maximum at 07:05\,UT.

The radio counterpart of this flare was a~22 minutes long radio event lasting from the
06:50 to 07:12\,UT recorded by the \textit{Giant Metrewave Radio Telescope} (GMRT)
in Pune, India. The \textit{Michelson Doppler Imager} (MDI, \citeauthor{Scherrer95} \citeyear{Scherrer95}) onboard the SoHO
spacecraft \cite{Domingo95} performed routine full-disk observations with a~96~minute
cadence. The magnetogram nearest to the radio event was observed at 06:24\,UT. The EUV
counterparts of this flare were observed by the \textit{Extreme Ultraviolet Imaging
Telescope} \cite{Delaboudiniere95} onboard the SoHO spacecraft.

%
%
\begin{figure}[]
\begin{center}
\includegraphics[scale=0.50]{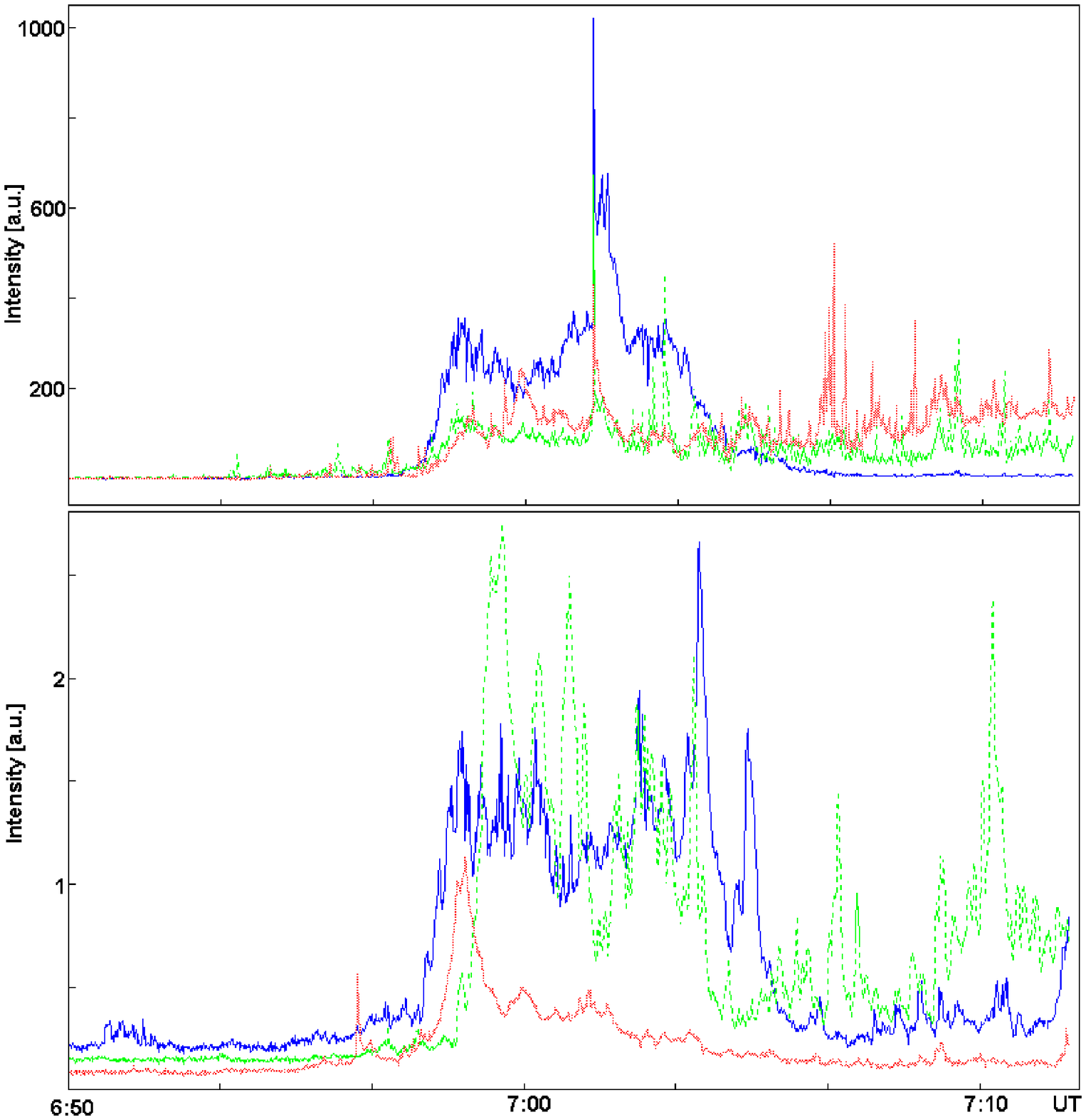}
\caption{Light curves of the 26~November~2005 radio event obtained from the GMRT maps.
         Upper panel: Radio fluxes at 244\,MHz corresponding to the sources U1~(blue),
         U2~(dashed green) and U3~(dotted red).
         Bottom panel: Radio fluxes at 611\,MHz corresponding to the sources D1~(blue),
         D2~(dashed green) and D3~(dotted red).}
\label{figure3}
\end{center}
\end{figure}

%
%
\begin{figure}[]
\begin{center}
\includegraphics[scale=0.8]{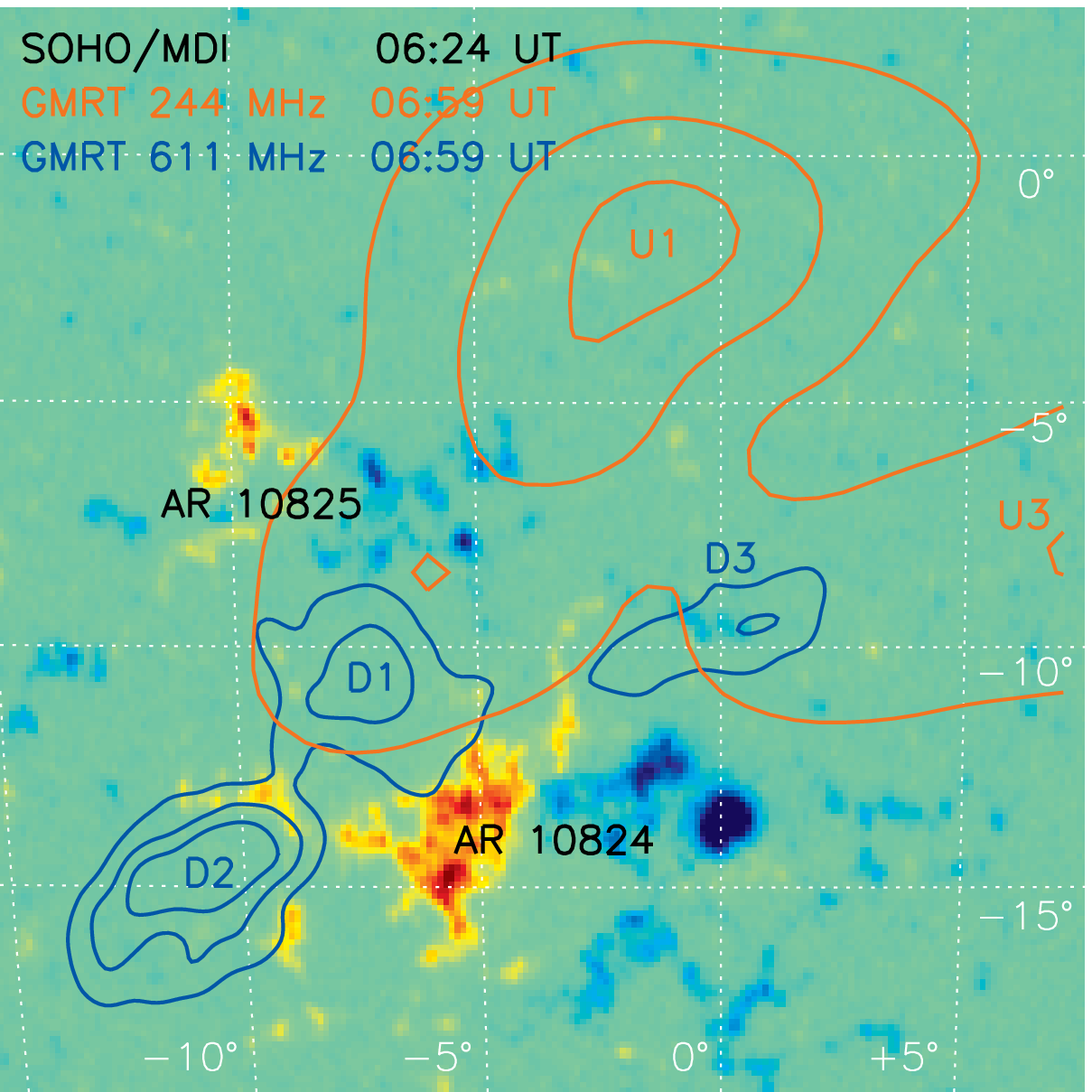}
\caption{Superposition of the radio sources at 244\,MHz (orange contours)
         and 611\,MHz (blue contours) observed at 06:59\,UT on top of the
         photospheric magnetic
         map obtained with the SoHO/MDI at 6:24\,UT. Individual radio sources
         are labelled. The MDI magnetogram is saturated to $\pm$\,10$^3$\,G.
         Heliographic coordinates are shown for orientation. Solar North is up.}
\label{figure4}
\end{center}
\end{figure}

\subsection{GMRT observations and data analysis}
\label{Sect:2.1}

On 26~November~2005, the GMRT observed the Sun at two
frequencies, 244 and 611\,MHz. The GMRT instrument \cite{Swarup91,Ana02,Mercier06} is
a~radio interferometer consisting of 30 fully steerable individual radio telescopes. Each
telescope has a~parabolic antenna with a~diameter of 45\,m and 16 of these individual
antennas are arranged in an Y~shape array with each arm having length of 14\,km from the
array centre. The remaining 14 telescopes are located in the central area of 1\,km$^2$.
The interferometer operates at wavelengths longer than 21 cm, with six frequency bands
centered on the 38, 153, 233, 327, 610, and 1420\,MHz frequencies. The maximum resolution
depends on the configuration, and varies between 2~and 60\,arc\,secs.

The observed interferometric data at 244 and 611\,MHz were Fourier transformed to
generate a~series of 1320 snapshot images of the Sun at 1~second time-cadence, from 06:50
to 07:12\,UT. The images were cleaned using the algorithm developed by
\inlinecite{Schwab84} and rotated to correct for the solar North. The synthesized beam
dimensions giving the GMRT positional error are are 77.7~$\times$~50.8 and
17.7~$\times$~13.4\,arc\,sec at 244 and 611\,MHz, respectively.

An example of the observations showing the main radio sources is shown in Figure
\ref{figure1}. On this figure, six sources can be identified, three for each frequency.
At the frequency of 244\,MHz, the sources are outlined in the left panel and denoted as
U1, U2, and U3. The main sources at the frequency 611\,MHz are shown in the right panel
of Figure \ref{figure1} and labelled as D1, D2, and D3. The evolution of these sources
during the radio event is shown in Figure \ref{figure2}. Positions of the individual
radio sources U1--U3 and D1--D3 are shown in red. Notable is the merging of the sources
U1 and U2 at around 6:59\,UT. The source U3 remains far from the others (U1 and U2).

We constructed light curves of the individual radio sources by enclosing these sources in
rectangular regions on the GMRT maps. The light curves are shown in Figure \ref{figure3}.
The radio fluxes at 244\,MHz corresponding to the sources U1, U2, and U3 are shown in the
upper panel as the blue, dashed green and dotted red lines, respectively. The radio
fluxes at 611\,MHz frequency corresponding to the sources D1~(blue), D2~(dashed green)
and D3~(dotted red) are present in the bottom panel. The calibration of the individual
fluxes is relative only and is expressed in arbitrary units (a.u.).

Analysis of these time series of radio fluxes shows different time evolution profiles for
the different radio sources. The sources U1, D1, and D3 show about 5~minutes durations
with a~well defined main peak. On the other hand, the sources U2, U3, and D2 show
profiles with a~gradual rise and several peaks on top of a~roughly constant background.
The temporal properties of these sources are presented in Table~1. The start and end
times of each source were determined as the times when the fluxes were above or below
half the average flux of the whole burst profile of the source, respectively.

The cross-correlation coefficients between pairs of all GMRT U~and~D sources show that
there is only one pair with correlation higher than~75\%. This pair is the sources U1 and
D1 with the cross-correlation coefficient of~81\%. This degree of correlation indicates
that these radio sources can have a~common origin.

The maximum of the cross-correlation coefficient is flat if the time lag of U1 with
respect to D1 ranges 0--50\,s. It indicates that first parts of radio fluxes D1 and U1
are correlated by some fast moving agents (possibly beams) and the second parts by slowly
moving agents (possibly waves).

%
%
\begin{figure}[]
\begin{center}
\includegraphics[scale=0.9]{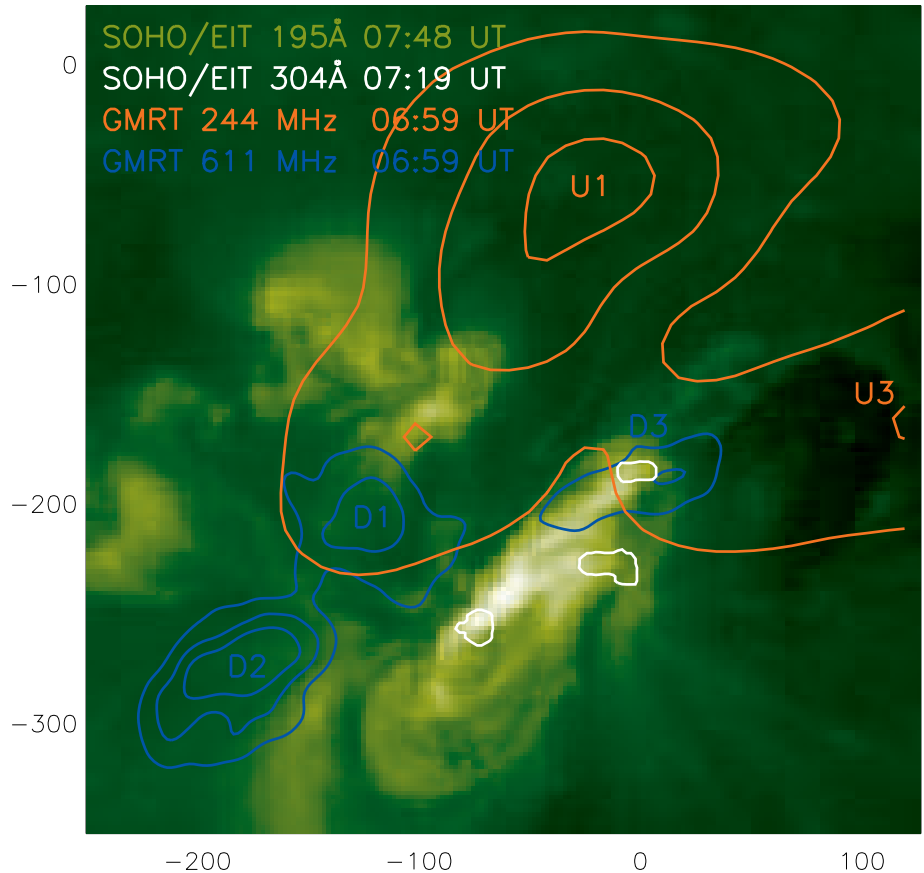}
\caption{Superposition of the radio sources (same as in Figure \ref{figure4}) on top of the
         SoHO/EIT 195\,\AA~observations taken at 7:48\,UT. The
         EIT 195\,\AA~observations are saturated to 2$\times$10$^3$ DN\,s$^{-1}$\,px$^{-1}$
         and shown in logarithmic intensity scale for better visibility of fainter
         emitting loop systems. The flare arcade footpoints observed using the
         filter 304\,\AA~at 07:19\,UT are shown as white contours corresponding
         to 10$^3$ DN\,s$^{-1}$\,px$^{-1}$.}
\label{figure5}
\end{center}
\end{figure}

\subsection{Photospheric magnetic field and the EUV flare}
\label{Sect:2.3}

The GMRT interferometric observations offer the advantage of direct spatial comparison
with the photospheric magnetic field and the EUV flare morphology, as observed by the
SoHO/MDI and SoHO/EIT instruments, respectively. The comparison of the radio sources loci
with the MDI magnetogram is presented in Figure \ref{figure4}. The magnetogram was
observed at 06:24\,UT and rotated to the time 06:59 corresponding to the radio
observations using the SolarSoftware routine {\it drot\_map.pro}. The time of the radio
observations, 06:59\,UT, is chosen because it corresponds to the times of the first radio
peaks in the sources U1 and D1.

The radio sources are located in a~quadrupolar magnetic configuration consisting of two
active regions, NOAA 10824 and 10825 (Figure \ref{figure4}). Both active regions are
bipolar with a~$\beta$-configuration. The AR 10824 is located approximately at the
heliographic latitude of $\approx -13^\circ$ and contains a~well-developed, leading
negative-polarity sunspot. Other polarities consist of plage regions or pores, which is
the case also for the AR 10825 located at latitudes of $\approx -7^\circ$. A~notable
feature is that the radio sources are {\it not} located on top of the main magnetic
polarities, but, in general, they overlie weak-field regions.

The SoHO/EIT instrument was performing full-disc observations in the 195\,\AA~filter with
a~cadence of approximately 12 minutes. Observations in other filters, \textit{i.e.} 171,
284 and 304\,\AA~were performed at 07:00, 07:06 and 07:19\,UT, respectively. Due to poor
pointing information of the EIT instrument, the EIT 304\,\AA~observations were coaligned
manually with the MDI observations by matching the EIT 304\,\AA~brightenings with small
magnetic polarities observed by MDI, which show good spatial correlation
\cite{Ravindra03}. We estimate the error of this manual coalignment to be
$\approx$~5\,arc\,sec.

In the EIT 304\,\AA~observations, three flare arcade footpoints are discernible, shown in
Figure \ref{figure5} as white contours corresponding to the observed intensity of 10$^3$
DN\,s$^{-1}$\,px$^{-1}$. All three footpoints are located well within the AR 10824, with
one of them in the positive polarity and the other two in the negative polarities North
of the sunspot. EIT 195\,\AA~observations show cooling system of flare loops connecting
these three footpoints. The flare loops are well-visible at 195\,\AA~at 07:48\,UT and are
shown in Figure \ref{figure5}. The global magnetic configuration of AR~10824 is that
of a~sigmoid with large shear. The magnetic configuration of the neighboring AR 10825 is
near-potential.

Comparison of the loci of the radio sources with the EUV flare morphology (Figure
\ref{figure5}) shows that the radio sources have no direct spatial correspondence with
the EUV flare loops or their footpoints. The source D3 is an exception, since it overlies
a~portion of the flare loops. The sources U1, U3, D1 and D2 are located in the area of
weak EUV emission.

%
%
\begin{table}[]
\caption[]{Basic parameters derived for radio bursts at GMRT frequencies.}
\label{table2}
\begin{tabular}{cccc|c}
\hline
\hline
Frequency   & Source    & Fundamental        &  Plasma density  &    First harmonic    \\
            &           & frequency altitude &                  &    altitude          \\
 $[$MHz]    &           &   [Mm]             &   [cm$^{-3}$]    &      [Mm]            \\
\hline
 244        & U1--U3    &      48            & 7.4 x 10$^{8}$   &         86            \\
 611        & D1--D3    &      22            & 4.6 x 10$^{9}$   &         40            \\
\hline
\end{tabular}
\end{table}

%
%
\section{Magnetic structure of active regions NOAA 10824 and 10825}
\label{Sect:3}

\subsection{Magnetic field extrapolation and the altitude of the radio emission}
\label{Sect:3.1}

\begin{figure}[]
\begin{center}
\includegraphics[scale=1.0]{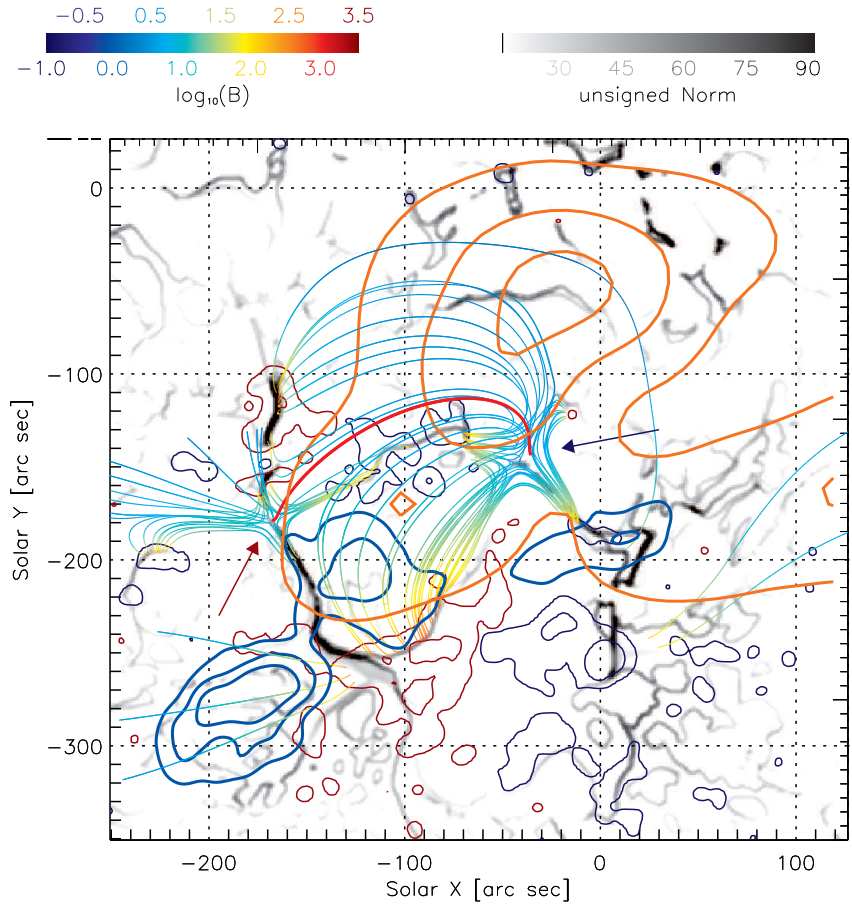}
\caption{Plane-of-the-sky projection of the extrapolated magnetic field.
Red (SE) and blue (NW) arrows show the locations of the positive and negative coronal magnetic null
points, respectively. The separator lying at the intersection of the respective fans is
plotted as thick red line. The magnetic field lines are colored according to the local
magnetic field, in the range of log$_{10}(B/\mathrm{T})$\,$\in$\,$\left<-1.0,3.5\right>$.
The thin, dark-red and dark-blue contours denote positive and negative photospheric
polarities, respectively. The two contour levels correspond to $B_z(z=0) = \pm$ 50 and
500\,G. Shades of gray show the photospheric intersections of the quasi-separatrix
layers. The contours of the GMRT sources are located at the altitudes corresponding to
the fundamental frequency (Table~\ref{table2}) and are shown as thick orange (U1--3) and
blue (D1--3) contours, respectively.}
\label{figure6}
\end{center}
\end{figure}

\begin{figure}[]
\begin{center}
\includegraphics[scale=1.0]{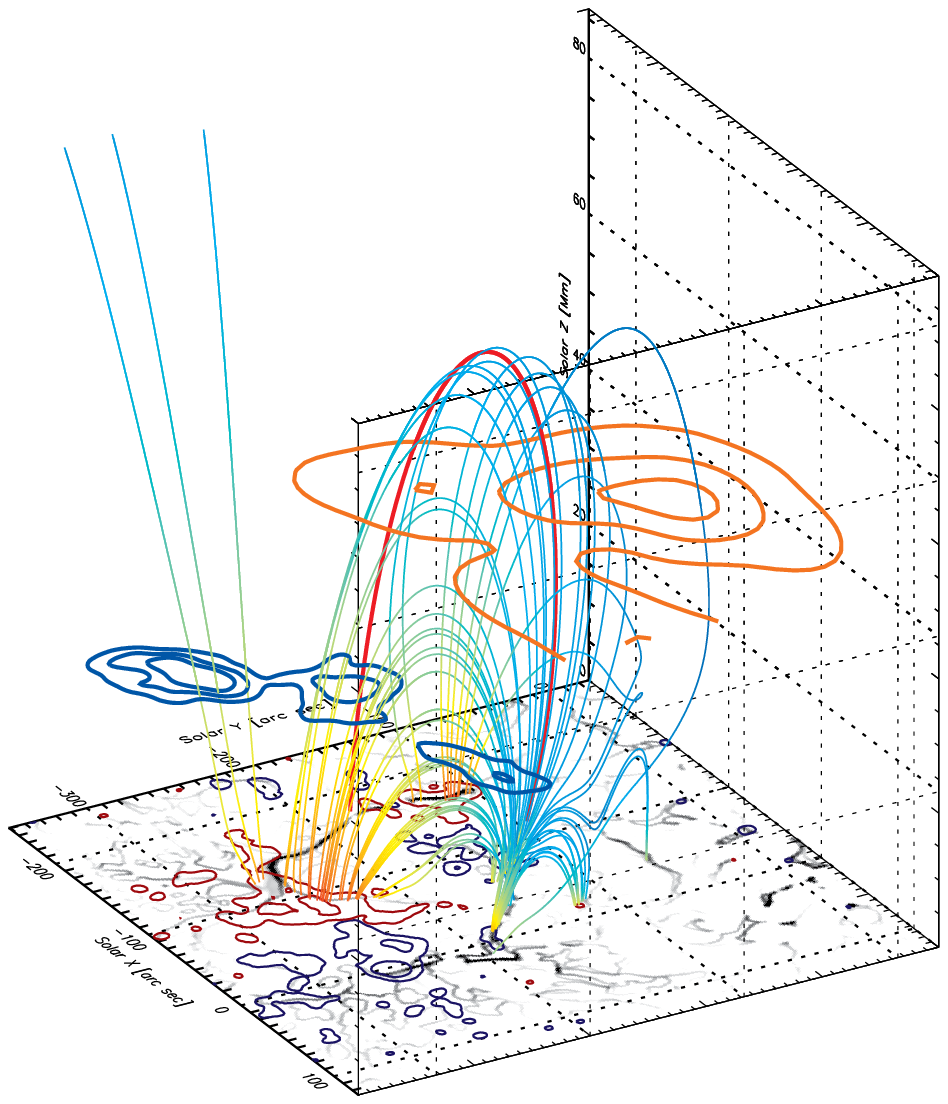}
\caption{Same as in Figure~\ref{figure6}, but in a~3D~projection to show the
         altitudes of the individual field lines as well as individual radio
         sources. Only the field lines lying in the vicinity of the negative
         null point or passing through the source D2 are shown for clarity.}
\label{figure7}
\end{center}
\end{figure}

To investigate the relationship between the radio sources and the structure of the
magnetic field of active regions NOAA 10824 and 10825, we performed an extrapolation of
the SoHO/MDI magnetogram (Figure \ref{figure4}) observed at 06:24\,UT prior to the flare
and associated radio events. The extrapolation was carried out in linear force-free
approximation, where the magnetic field $\vec{B}$ given by the solution of the equation
\begin{equation}
    \vec{\nabla} \times \vec{B} = \alpha\vec{B}\,,
    \label{potential}
\end{equation}
for $\alpha = const$. The solution is subject to the boundary condition $B_z(x,y,z=0)$
given by the observed magnetogram, where $0 < x < L_x$ and $0 < y < L_y$. The constant
$\alpha$ is subject to the condition $\alpha < \alpha_\mathrm{max} = 2\pi /
\mathrm{max}(L_x, L_y)$, otherwise the magnetic field is non-physical. We utilized the
Fourier transform method developed by \inlinecite{Alissandrakis81} and
\inlinecite{Gary89}. This method allows for extrapolation of the part of the observed
magnetogram in a~carthesian geometry. The computational box is shown in Figures
\ref{figure6} and~\ref{figure7}.

We calculated a~range of linear force-free models with various values of~$\alpha$.
However, the flare loops are poorly approximated with $\alpha = const.$, even with large
values of $\alpha$ close to $\alpha_\mathrm{max}$. The reason for this probably is the
presence of differential shear within the active region \cite{Schmieder96}. Moreover,
using large values of $\alpha$ leads to poor fit to the observed shape of the coronal
loops in the AR 10825, which is close to the potential state ($\alpha = 0$). Therefore,
we chose to extrapolate in the potential approximation. We also note that the potential
approximation does usually a~good job in capturing the topological structure of the
active region (though not at sigmoid locations, \textit{e.g.} \citeauthor{Schmieder03}
\citeyear{Schmieder03}).

To compare the 3D~magnetic field geometry with the loci of the radio sources, observed in
a~2D~plane of the sky, the approximate altitude at which the radio emission originates
must be determined. To do that, we consider the parameters of the radio bursts (at GMRT
frequencies) and the solar atmosphere density model \cite{Aschwanden02} for the radio
plasma emission at fundamental frequency and the first harmonic (see also
Section~\ref{Sect:3.2}). We use Aschwanden's density model because it was derived
from radio observations. The basic parameters of the radio sources are derived and
summarized in Table~\ref{table2} where the plasma density values belong to the
fundamental frequency altitude.

%
%

%
%

\subsection{Magnetic topology and its relation to radio sources}
\label{Sect:3.2}

The extrapolated magnetic field contains a~pair of coronal magnetic null points between
the two active regions. The negative and positive coronal null points are denoted
by red (SE) and blue (NW) arrows in Figure \ref{figure6}. On this figure, the dark-red and
dark-blue contours correspond to positive and negative photospheric polarities,
respectively, with $B_z$\,=\,$\pm$50\,and\,500\,G. Since the magnetic field of
AR~10825 is weaker, the null points are located closer to this active region than to the
larger AR 10824. The existence of these null points is a~consequence of the multipolar
structure of the magnetic field (\textit{e.g.} \citeauthor{Longcope05}
\citeyear{Longcope05} and references therein). The magnetic structure in the vicinity of
these null points is shown using several field lines regularly distributed
near the null points. The fan surfaces of these null points intersect, forming
a~separator \cite{Baum80,Longcope05}. The separator is found manually by trial-and-error
and is shown by a~thick red line.

Both the spine and the fan surface of the negative null point are closed within
the computational domain, while the fan surface of the positive null point is
open. This is not typical of coronal null points \cite{Pariat09,DelZanna11}.
Both fan surfaces form a~part of the quasi-separatrix layers
\cite{Priest95,Demoulin96,Demoulin97} associated with the active regions. The
intersections of these quasi-separatrix layers with the photosphere are shown
in Figures \ref{figure6} and~\ref{figure7} by shades of gray corresponding to
the unsigned norm~$N$ of the field line footpoint mapping, defined as
\begin{equation}
    N(x,y)  = \left( \left(\frac{dX}{dx}\right)^2 +\left(\frac{dX}{dy}\right)^2 +\left(\frac{dY}{dx}\right)^2 +\left(\frac{dY}{dy}\right)^2 \right)^{1/2}\,,
    \label{norm}
\end{equation}
where the starting footpoint of a~field line at $z~=~0$ is denoted as $(x,y)$ and its
opposite footpoint as $(X,Y)$. The unsigned norm is computed irrespective of the
direction of the footpoint mapping. We note that the correct definition of
quasi-separatrix layers is through the squashing factor $Q$ or the expansion-contraction
factor $K$ \cite{Titov02}, which are invariants with respect to the footpoint mapping
direction. Here, we use the unsigned norm, since it is much less sensitive to the errors
arising because of the measurement noise and numerical derivation due to finite
magnetogram resolution, and at the same time gives a~good indication of the
expansion-contraction factor~$K$.

The sources~U and~D are plotted as thick blue and thick orange contours, respectively.
These sources are plotted in height corresponding to the altitudes given by the
fundamental frequencies (Table~\ref{table2}). It is clearly seen that both U1 and D1
are lying along the fan of the negative null point, with the field lines passing through the
D1 being separated by the separator (Figures \ref{figure6} and~\ref{figure7}). Therefore,
both the U1 and D1 lie along a~common magnetic structure, providing straightforward
explanation for the observed correlation between these two sources
(Section~\ref{Sect:2.1}). The separator also passes through U1, as does some other fan
field lines rooted in AR~10825.

We note that the fan of the negative null-point covers only a~part of the source U1. The
rest of the U1 is associated with closed field lines rooted in the positive polarity of
the AR 10825, in direct vicinity of the footpoints of the fan field lines. One example of
such field lines is plotted in Figures~\ref{figure6} and~\ref{figure7}. That there are
structures other than the fan of the negative null-point passing through the source U1
offers explanation for the fact that the correlation coefficient between U1 and D1 is
smaller than one.

The sources D2 and U3 are located on the field lines that are open within the
computational box. The footpoints of these field lines are located in the AR~10824 in the
vicinity of the photospheric quasi-separatrices with highest $N$ separating the open and
closed magnetic flux (Figure \ref{figure6}). Overall, we conclude that the field lines
passing through the sources D1, D2, U1 and U3 are rooted in the vicinity of the flare
arcade footpoints (compare Figures \ref{figure5} and~\ref{figure6}), while the source D3
appears to be connected directly to flare loops. These observations point out the fact
that even a~localized, weak flare can cause a~widespread perturbation involving a~large
portion of the surrounding magnetic field and its topological features.

We stress that the goodness of the match between the extrapolated magnetic field and the
radio sources is subject to the approximations used. \textit{E.g.} presence of electric
current in the model of the magnetic field would alter the geometry of the magnetic
field. In particular, using a~linear force-free approximation would mostly affect the
longest field lines passing through the source U1. Additionally, the density model of
\cite{Aschwanden02} provides only a~single altitude for the emission of each radio
frequency, while in reality the density profile and thus the altitude of emission can be
different for each field line. However, the available longitudinal MDI magnetogram and
the EIT data do not provide necessary constraints (Section~\ref{Sect:2}) for using more
elaborate models of the magnetic field or density throughout the observed radio-emitting
atmosphere.

We also note that the magnetic connectivity between U1 and D1 exists only if these
sources are assumed to emit at the fundamental frequency (Table~\ref{table2}). If the
origin of the radio emission would be the second harmonic, the radio emission should
occur at altitudes which are higher than the vertical extent of the negative null point
fan. This point is discussed further in Section~\ref{Sect:5}.


%
%
\begin{table}[]
\caption[]{Characteristics periods~$P$ for wavelet power spectra showing a~tadpole pattern.}
\label{table3}
\begin{tabular}{ccc}
\hline
\hline
Source & Frequency & Period $P$                 \\
       &  $[$MHz]  &   [s]                      \\
\hline
U1     &   244     & 83, 25, 22, 11, 10         \\
D1     &   611     & 76, 22, 13                 \\
\hline
\end{tabular}
\end{table}

%
%
\begin{figure}[]
\begin{center}
\includegraphics[scale=0.60]{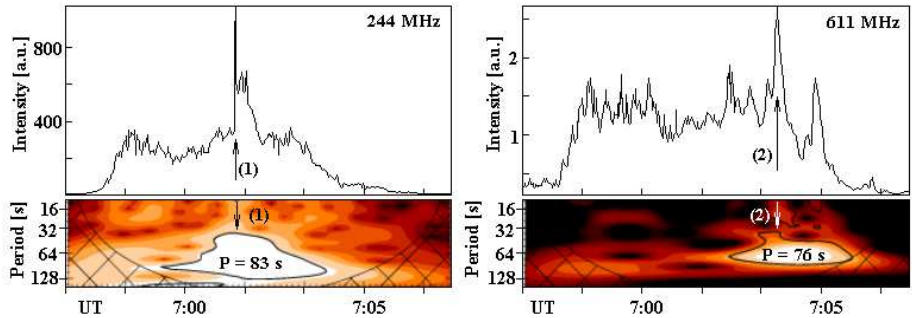}
\caption{Examples of the wavelet power spectra showing tadpole patterns with
         characteristics periods~$P$ (bottom panels) in comparison with the GMRT radio
         fluxes (upper panels) at 244\,MHz (source U1) and 611\,MHz (source D1) frequencies.
         The lighter area shows a~greater power in the power spectrum and it is bound
         by the solid contour of the confidence level~$>$95\%. The hatched regions belong
         to the cone of influence with edge effects due to finite-length time series.
         The arrows (1) and (2) correspond to 7:01:30 and 7:03:52\,UT, respectively.}
\label{figure8}
\end{center}
\end{figure}

\begin{figure}[]
\begin{center}
\includegraphics[scale=0.4]{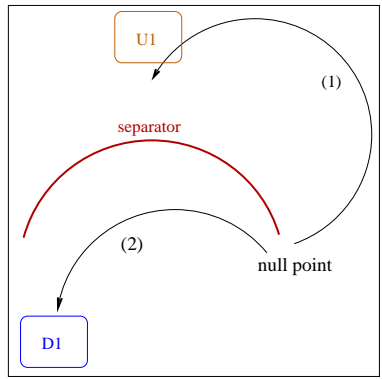}
\caption{Scheme of the suggested scenario for the well-correlated sources U1 and D1
         lying along the fan of the coronal null point. Magnetoacoustic waves move along
         magnetic field lines (trajectories (1) and (2)). The separator is located between
         these trajectories.}
\label{figure9}
\end{center}
\end{figure}

%
%
\section{Wavelet analysis of the GMRT time series}
\label{Sect:4}

To investigate the nature of the disturbance giving rise to the radio sources, an
analysis of possible periodicity of radio time series at all frequencies, detected by the
GMRT instrument, was made using the Morlet wavelet transform method \cite{Torrence98}. We
focused on the well correlated GMRT sources U1 and~D1.

The wavelet power spectra at all frequencies of the GMRT instrument show the tadpole
patterns \cite{Nakariakov04} with characteristics periods~$P$
(\citeauthor{Meszarosova09a}, \citeyear{Meszarosova09a, Meszarosova09b}) in the range
10--83\,s.

These periods were detected in the entire studied time interval 06:50--07:12\,UT and are
listed in Table~\ref{table3}. The periods of about 10\,s were found at all frequencies.
Figure~\ref{figure8} shows two examples of the wavelet power spectra of the well
correlated radio sources U1 and D1 showing tadpole patterns (bottom panels) with the
longest characteristic periods $P$ listed in Table~\ref{table3}. In this figure, the
arrows (1) and (2) are used to denote the times of 7:01:30 and 7:03:52\,UT, respectively.
These times correspond to the main peaks of the flux shown in the upper panels and the
occurring tadpole head maximum shown in bottom panels. It has been proposed that the tadpoles indicate
arrival of the magnetoacoustic wave trains at the radio source
(\citeauthor{Nakariakov04}, \citeyear{Nakariakov04}; \citeauthor{Meszarosova09a},
\citeyear{Meszarosova09a, Meszarosova09b}).

Although the U1 and D1 radio fluxes are well correlated (81\%) the sources U1 and D1 are
not the same in all details. Different appearance of the tadpoles in Figure~\ref{figure8}
reflects different plasma parameters in the sources U1 and D1 \cite{Jelinek12}. This is
expected, since these tadpoles have different characteristic period~$P$, which means that
they are propagating in different waveguides (see Section~\ref{Sect:5}).

%
%
\section{Discussion and Conclusions}
\label{Sect:5}

We studied solar interferometric maps (Figures \ref{figure1} and~\ref{figure2}) and time
series of the six individual radio sources, U1--U3 at 244\,MHz and D1--D3 at 611\,MHz
(Figure~\ref{figure3}), observed by the GMRT instrument with 1\,s time resolution during
the B8.9~flare on 26~November~2005. We determined that only sources U1 and D1 are well
correlated with the cross-correlation coefficient of~81\%.

Generally, the positions of the radio sources can be influenced by refractive
effects. They play a~role in the solar atmosphere as well as in the Earth ionosphere. As
concerns to the effects in the solar atmosphere, our low frequency 244\,MHz source U1
(which is the main feature under study together with the source D1) is nearly in the disk
center, see Figure~\ref{figure1} or~\ref{figure4}. In the disk center the refractive
effects come to zero. Similarly, the D1 source at 611\,MHz is close to the disk center.
We note that the phase calibration process at GMRT corrects the ionospheric refraction.

Comparison of the GMRT interferometric maps with the EUV observations of the flare by the
SoHO/EIT instrument showed that only the source D3 can be connected to the system of
flare loops. All other radio sources are located well away from the X-ray and EUV flare
(\textit{cf.} \citeauthor{Benz01} \citeyear{Benz01}). Potential extrapolation of the
observed SoHO/MDI magnetogram in combination with the solar coronal density model of
\inlinecite{Aschwanden02} allowed us to find the connection of these other radio sources
to the magnetic field of the observed active regions and their magnetic topology. The
sources U1 and D1 were found to lie along the fan of a~negative coronal null point. That
the sources U1 and D1 lie along a~common magnetic structure offers explanation for their
observed correlation. The sources D2 and U3 lie at open field lines anchored near the
quasi-separatrices in the vicinity of the flare arcade footpoints.

The connection between the radio sources and the magnetic field, as found in Section
\ref{Sect:3.2} is valid only if the radio emission originates at the fundamental
frequency. This is because the altitudes corresponding to the first harmonic are too high
(Table~\ref{table2}). We note that the emission at the first harmonic is usually
considered to be stronger than at the fundamental frequency, due to the strong absorption
for the fundamental emission. However, there are cases (\textit{e.g.} enhanced plasma
turbulence in the radio sources) which reduces this absorption and thus the emission at
the fundamental frequency can be stronger than at the first harmonic.

Based on the results above, we conclude that even the observed localized flare is able to
cause a~widespread perturbation involving a~large portion of the topological structure of
the two active regions. This perturbation can be understood in the following terms: As
the flare progresses, the flare arcade footpoints move away from each other. Since these
footpoints correspond to the intersections of the quasi-separatrix layers with the
photosphere \cite{Demoulin96,Demoulin97}, the increase of their distance perturbs the
surrounding magnetic field and excite waves. If the surrounding, perturbed magnetic field
contains null points, these can collapse and form a~current sheet, thereby commencing
reconnection in regions not directly involved in the original flare. This can then be
also the mechanism for sympathetic flares and eruptions
\cite{Moon02,Khan06,Jiang08,Torok11,Shen12}.

For the first time, wavelet tadpole patterns with the characteristic periods of 10--83\,s
(Table~\ref{table3}) were found at metric radio frequencies. We have interpreted them in
accordance with the works of \inlinecite{Nakariakov04} and \citeauthor{Meszarosova09a}
(\citeyear{Meszarosova09a,Meszarosova09b}) as signatures of the fast magnetoacoustic
waves propagating from their initiation site to studied radio sources U1 and D1.

The mechanism for initiation of the inferred magnetoacoustic waves present in the radio
sources can thus be both the observed EUV flare and the flare-induced collapse of the
null point located in the surrounding magnetic field. These waves propagated towards
radio sources along magnetic field lines. They arrived at the radio sources (see the
wavelet tadpoles in Figure~\ref{figure8}) and modulated the radio emission there. We
expect that the magnetoacoustic waves, through their density and magnetic field
variations, modulate the growth rates of instabilities (like the bump-on-tail or
loss-cone instabilities), which generate plasma- as well as the observed radio waves.

The schematic scenario for the well-correlated sources U1 and D1 lying along the fan of the
coronal null point is shown in Figure~\ref{figure9}. Magnetoacoustic wave trains move
along magnetic field lines (trajectories~1 and~2). The separator is located between these
trajectories (\textit{i.e.} between two groups of magnetic field lines (\textit{cf.}
Figure~\ref{figure6}). The distances between the null point and radio sources along the
magnetic field lines are obtained directly from the extrapolation and allow the
determination of the velocities of the magnetoacoustic waves. The averaged distances
between the radio source and the null point, are about 53 and 103\,Mm for the U1 and D1,
respectively. We considered the time intervals $t2-t1$ and $t3-t1$ where $t1$ is the time
of triggering of the magnetoacoustic wave trains. We assume that these trains propagating
to the source U1 as well as to the source D1 were generated simultaneously at the
beginning of the radio event, \textit{i.e.} at the start time of U1 and D1 sources
(Table~\ref{table1}). The times $t2$ and $t3$ correspond to the times of tadpole head
maxima (arrows (1) and (2), Figure~\ref{figure8}), respectively. Thus, the mean velocities of
the magnetoacoustic waves are 260 and 300\,km\,s$^{-1}$ for U1 and D1 sources,
respectively.

Now, let us compare these mean velocities with the Alfv\'en velocities at the radio
sources at U1 and D1. Taking plasma densities from Table~\ref{table2} and the magnetic
field from the extrapolation ($B$~=~3\,G at U1 and 13\,G at D1) the Alfv\'en velocity at
these sources are $v_A$~=~220 and 390\,km\,s$^{-1}$, respectively. These velocities are
in agreement with previous results, considering that the Alfv\'en velocities change along
the trajectory of these magnetoacoustic waves, starting from the coronal null point.

Thus, knowing the Alfv\'en velocities in the radio sources, we can estimate also the
width~$w$ of the structure through which the magnetoacoustic wave trains are guided.
Namely, the period of this magnetoacoustic wave can be estimated as $P~\simeq~w/v_A$
(Nakariakov \etal, 2004). Thus, the widths~$w$ of the structures guiding the
magnetoacoustic waves (modulating the radio emission) are 18\,Mm
(83\,s~$\times$~220\,km\,s$^{-1}$) and 30\,Mm (76\,s~$\times$~390\,km\,s$^{-1}$) in the
U1 and D1 radio sources, respectively. This rough estimation confirms that the extent of
the structure guiding magnetoacoustic waves is within the width of the fan of magnetic
field lines in both radio sources. The wavelet tadpoles with shorter periods
(Table~\ref{table3}) show that within the large structure guiding long period
magnetoacoustic wave trains ($P\sim$~80\,s), there are also narrower waveguides guiding
shorter period waves.

The complex of radio sources U1--U3 and D1--D3 includes the whole topological structure
of the active area contrary to the rather localized EUV sources (Figure~\ref{figure5}).
This is in line with the observed radio sources being usually larger that the real
sources due to scattering of radio waves (\textit{e.g.} \citeauthor{Benz02},
\citeyear{Benz02}).

Considering the above presented estimations we can conclude that they support the
scenario (Figure~\ref{figure9}) of magnetoacoustic waves (wavelet tadpoles) in a~fan
structure above the coronal magnetic null point.

\acknowledgements We are thankful to Dr. E.~Dzif\v{c}\'akov\'a for encouraging
discussions. H.~M. thanks Dr.~J.~Ryb\'ak for his help with the wavelet analysis that was
performed using the software based on tools provided by C.~Torrence and G.~P.~Compo at
\texttt{http://paos.colorado.edu/research/wavelets}. H.~M. acknowledges support from
the PCI/INPE Grant 33/2011 of the National Space Research Institute in Brazil. H.~M. and
M.~K. acknowledge support from the Grant GACR P209/12/0103, the research project
RVO:67985815 of the Astronomical Institute AS and the Marie Curie PIRSES-GA-2011-295272
RadioSun project. J.~D. and M.~K. acknowledge support from the bilateral project No.
SK-CZ-11-0153. J.~D. also acknowledges the support from the Scientific Grant Agency,
VEGA, Slovakia, Grant No. 1/0240/11, Grant No. P209/12/1652 of the Grant Agency of the
Czech Republic, Collaborative grant SK-CZ-0153-11 and Comenius University Grant
UK/11/2012. We also thank the staff of the GMRT who have made these observations possible
as well as the SoHO/MDI and SoHO/EIT consortia for their data. GMRT is run by the
National Centre for Radio Astrophysics of the Tata Institute of Fundamental Research.

\bibliographystyle{spr-mp-sola-cnd}
\bibliography{gmrt}

\end{article}
\end{document}